\documentclass[11pt, fceqn]{article}
 \usepackage{amsfonts}
 \usepackage{flafter}
 \usepackage{amssymb,graphics,graphicx,subfigure,caption2,rotating}
 \usepackage{amsmath, latexsym}
\usepackage[amsmath,thmmarks]{ntheorem}
 \usepackage[top=3.5cm]{geometry}
\begin{document}
\date{}
\title{Entanglement detection via some classes of measurements}
\author{Shu-Qian Shen$^1$\thanks{E-mail: sqshen@upc.edu.cn.}, Ming Li$^1$\thanks{E-mail: liming@upc.edu.cn. }, Xue-Feng Duan$^2$\thanks{E-mail:duanxuefenghd@aliyun.com.}\\
{\small{\it $^1$College of Science, China University of Petroleum, Qingdao, 266580, P.R. China}}\\
{\small{\it $^2$College of Mathematics and Computational Science,}}\\
{\small{\it Guilin University of Electronic Technology, Guilin, 541004, P.R. China}}}\maketitle
\begin{abstract}
Based on the mutually unbiased bases, the mutually unbiased measurements and the general symmetric informationally complete positive-operator-valued measures, we propose three separability criteria for $d$-dimensional bipartite quantum systems, which are more powerful than the corresponding ones introduced in [C. Spengler, M. Huber, S. Brierley, T. Adaktylos, and B.C. Hiesmayr, Phys. Rev. A \textbf{86}, 022311 (2012); B. Chen, T. Ma, and S.M. Fei, Phys. Rev. A \textbf{89}, 064302 (2014); B. Chen, T. Li, and S.M. Fei, arXiv:1406.7820v1 [quant-ph] (2014)]. Some states such as Werner states and Bell-diagonal states are used to further illustrate the efficiency of the presented criteria.
%
%
\end{abstract}

\section{Introduction}
Entanglement regarded as a physical resource plays a key role in the quantum information and computation \cite{Nielsen2010}. Naturally, one of the fundamental problems is determining whether a given quantum state is entangled or not. This problem is easy for pure states; see, e.g., \cite{Albeverio2003}. However, until now, there are no general practical criteria for general mixed states \cite{Gurvits2003-1}. In the last decades, many efforts have been made to give various special criteria for entanglement detection. There are PPT criteria \cite{ppt,Horodecki1997}, realignment criteria \cite{Albeverio2003,realignment,Zhang2008}, correlation matrix criteria \cite{correlation}, local uncertainty relation criteria \cite{uncertainty}, quantum Fisher information criteria \cite{Li2013} and so on;  see, e.g., \cite{survey} for a comprehensive survey.

Recently, some quantum measurements have been used to detect entanglement. In \cite{Spengler2012}, the authors first connected mutually unbiased bases (MUBs)\cite{Schwinger1960} with the detection of entanglement in bipartite, multipartite and continuous-variable quantum systems. For the bipartite case, the example of isotropic states showed that a complete set of MUBs can provide a more efficient criteria \cite{Spengler2012}. However, it is still not known that whether there exists a complete set of MUBs or not in non-prime-power dimensions \cite{Durt2010}. In \cite{Kalev2014}, the concept of MUBs is generalized to  mutually unbiased measurements (MUMs), which is then used to detect entanglement in the bipartite quantum systems \cite{Chen2014-1}. Unlike MUBs, for some special choices of the involved parameter, the complete set of MUMs can be constructed explicitly in all finite dimensions. Using similar ideas, Chen et al. \cite{Chen2014-2} proposed another criteria based on the general symmetric informationally complete (GSIC) positive-operator-valued measures (POVMs) \cite{Kalev2013}. Compared with the criteria based on MUMs, this criteria requires less local measurements.

  The criteria mentioned above have mainly two merits. Firstly,  they are powerful in the detection of entanglement \cite{Spengler2012,Chen2014-1,Chen2014-2}. Secondly, in comparison with other criteria including, for example, PPT criteria, realignment criteria and correlation matrix criteria, these criteria can be relatively easy to be implemented experimentally, since they only depend on some local measurements.

 In this paper, by making use of MUBs, MUMs and GSIC-POVMs, we propose three separability criteria based on $\rho-\rho^A\otimes \rho^B$, where $\rho$ is a bipartite density matrix in $\mathbb{C}^d\otimes \mathbb{C}^d$, $\rho^A$ ($\rho^B$) is the reduced density matrix of the first (second) subsystem. These derived criteria also provide easily experimental ways in the entanglement detection of unknown quantum states. Strict proofs show that these criteria are stronger than the corresponding ones in \cite{Spengler2012,Chen2014-1,Chen2014-2}.
Moreover, some well-known examples are supplemented to verify the efficiency of the presented results. It is worth noting that the realignment criteria introduced in \cite{Zhang2008} also depends on $\rho-\rho^A\otimes \rho^B$.

The remainder of the paper is arranged as follows. In Section 2, some preliminaries about MUBs, MUMs and GSIC-POVMs are reviewed. In Section 3, we derive three separability criteria based on MUBs, MUMs and GSIC-POVMs, and provide proofs and examples to illustrate the efficiency of the proposed criteria. Some concluding remarks are given in Section 4.

\section{MUBs, MUMs and GSIC-POVMs}
In $\mathbb{C}^d$, a set of orthonormal bases
\begin{equation}
\label{mubs}
\mathcal{D}_m=\{\mathcal{B}_1,\cdots,\mathcal{B}_m\} \text{ with }\mathcal{B}_k=\{|0_k\rangle,\cdots,|d-1_k\rangle\}
 \end{equation}
 is said to be a set of mutually unbiased bases (MUBs) if, for any $i\neq j$, $\mathcal{B}_i$ is mutually unbiased with $\mathcal{B}_j$, i.e., $|\langle l_i|m_j\rangle|^2=\frac{1}{d},l,m=0,\cdots,d-1$.
It has been shown in \cite{Wootters1989} that $m$ is at most $d+1$. Nevertheless, whether a complete set of MUBs, i.e., $m=d+1$, exists or not in non-prime-power dimensions is still unknown \cite{Durt2010} .

By the fact that an orthonormal base  $\mathcal{B}_k=\{|0_k\rangle,\cdots,|d-1_k\rangle\}$ can be equivalently defined as a positive-operator-valued measure (POVM) $\{E_0,\cdots,E_{d-1}\}$ with $E_i=|i_k\rangle\langle i_k|$, the concept of MUB is generalized to the mutually unbiased measurement (MUM) in \cite{Kalev2014} as follows. Two measurements on $\mathbb{C}^d$,
\[
\mathcal{P}^{(b)}=\left\{P_n^{(b)}|P_n^{(b)}\ge 0,\sum\limits_{n=1}^d P_n^{(b)}=I_d\right\}, b=1,2,
\]
 are said to be MUMs if and only if
\begin{itemize}
  \item Tr($P_n^{(b)})=1$;
  \item Tr$(P_n^{(b)}P_{ n'}^{(b')})=\delta_{nn'}\delta_{bb'} \kappa +(1-\delta_{nn'})\delta_{bb'}\frac{1-\kappa}{d-1}+(1-\delta_{bb'})\frac{1}{d}$,
\end{itemize}
where the efficiency parameter $\kappa$ satisfies $\frac{1}{d}<\kappa\le 1$. For the case $\kappa=1$, if $d+1$ MUMs in $\mathbb{C}^d$ exist, i.e., a complete set of MUMs exists, then the measurement operators reduce to rank one projectors given by MUBs. Thus, for a fixed parameter $\kappa$, the existence of a complete set of MUMs cannot be guaranteed. Nevertheless, for some special choices of $\kappa$, the complete set of MUMs can be explicitly constructed in all finite dimensions; see \cite{Kalev2014} for the concrete construction.

We now introduce the general symmetric informationally complete (GSIC) POVM established in \cite{Kalev2013}. The set $\left\{Q_m|Q_m\ge 0,\sum\nolimits_{m=1}^{d^2}Q_m=I_d\right\}$ on $\mathbb{C}^d$ is said to be a GSIC-POVM if and only if

\begin{itemize}
  \item Tr$(Q^2_m)=\alpha$;
  \item Tr$(Q_m Q_{ m'})=\frac{1-d\alpha}{d(d^2-1)},1\le m\neq m'\le d^2$,
\end{itemize}
where the parameter $\alpha$ satisfies $\frac{1}{d^3}<\alpha\le \frac{1}{d^2}$. From \cite{Kalev2013}, we have Tr($Q_m)=\frac{1}{d}$ for any $m$. For the case $\alpha=\frac{1}{d^2}$, if there exist GSIC-POVMs, then they reduce to SIC-POVMs; see, e.g., \cite{Scott2004} and references therein. Similar to MUMs, for some special choices of $\alpha$, the complete set of GSIC-POVMs can be constructed explicitly in all finite dimensions \cite{Kalev2013}.

\section{Entanglement detection}
\indent\indent\emph{Entanglement detection via MUBs.} Let $\mathcal{D}_m=\{\mathcal{B}_1,\cdots,\mathcal{B}_m\}$ be a set of MUBs defined as in (\ref{mubs}). It was shown in \cite{Spengler2012} that, if the state $\rho$ in $\mathbb{C}^d\otimes \mathbb{C}^d$ is separable, then
\begin{equation}
\label{emubs}
M_m(\rho)=\sum\limits_{k=1}^m\sum\limits_{i=0}^{d-1} \langle i_k|\otimes \langle i_k|\rho|i_k\rangle\otimes|i_k\rangle\le 1+\frac{m-1}{d}.
\end{equation}
In particular, for a complete set of MUBs, it holds
$
M_{d+1}(\rho)\le 2.
$
For simplicity, this criteria will be called MUB-criteria. We now present a separability criteria due to MUBs.\\
\\
\textbf{Theorem 1.} Let $\mathcal{D}_m=\{\mathcal{B}_1,\cdots,\mathcal{B}_m\}$ be a set of MUBs defined as in (\ref{mubs}). If the state $\rho$ in $\mathbb{C}^d\otimes\mathbb{C}^d$ is separable, then
\begin{align}
&L_m(\rho)=\sum\limits_{k=1}^{m}\sum\limits_{i=0}^{d-1}\left|\langle i_k|\otimes \langle i_k|\rho|i_k\rangle\otimes|i_k\rangle-\langle i_k|\rho^A|i_k\rangle \langle i_k|\rho ^B|i_k\rangle\right|\nonumber\\
&\label{memubs}\le \sqrt{\left(1+\frac{m-1}{d}-\sum\limits_{k=1}^{m}\sum\limits_{i=0}^{d-1}\langle i_k|\rho^A|i_k\rangle^2\right)\left(1+\frac{m-1}{d}-\sum\limits_{k=1}^{m}\sum\limits_{i=0}^{d-1}\langle i_k|\rho^B|i_k\rangle^2\right)}.
\end{align}
In particular, for a complete set of MUBs, it holds
\[
L_{d+1}(\rho)\le \sqrt{\left(2-\sum\limits_{k=1}^{d+1}\sum\limits_{i=0}^{d-1}\langle i_k|\rho^A|i_k\rangle^2\right)\left(2-\sum\limits_{k=1}^{d+1}\sum\limits_{i=0}^{d-1}\langle i_k|\rho^B|i_k\rangle^2\right)}.
\]
\textbf{Proof.} Any separable state $\rho$ can be written as $\rho=\sum\nolimits_{i=1}^{r}p_i\rho_i^A\otimes\rho_i^B$, where $0\le p_i\le 1,\sum\nolimits_{i=1}^{r}p_i=1$, $\rho_i^A(\rho_i^B)$ denotes the pure sate density matrix acting on the first (second) subsystem. Thus, it is easy to get
$\rho^A=\sum\nolimits_{i=1}^{r}p_i\rho_i^A, \rho^B=\sum\nolimits_{i=1}^{r}p_i\rho_i^B.$ By
\begin{equation}
\label{Zhang}
\rho-\rho^A\otimes\rho^B=\frac{1}{2}\left(\sum\limits_{s,t=1}^{r}p_sp_t(\rho_s^A-\rho_t^A)\otimes (\rho_s^B-\rho_t^B)\right)
\end{equation}
given in \cite{Zhang2008}, we obtain
\begin{align*}
L_m(\rho)&\le\frac{1}{2} \sum\limits_{k=1}^m\sum\limits_{i=0}^{d-1} \sum\limits_{s,t=1}^r \sqrt{p_sp_t}|\langle i_k|\rho_s^A-\rho_t^A|i_k\rangle|\sqrt{p_sp_t}|\langle i_k|\rho_s^B-\rho_t^B|i_k\rangle|\\
&\le\frac{1}{2} \sqrt{\sum\limits_{k=1}^m\sum\limits_{i=0}^{d-1} \sum\limits_{s,t=1}^rp_sp_t(\langle i_k|\rho_s^A-\rho_t^A|i_k\rangle)^2}\sqrt{\sum\limits_{k=1}^m\sum\limits_{i=0}^{d-1} \sum\limits_{s,t=1}^rp_sp_t(\langle i_k|\rho_s^B-\rho_t^B|i_k\rangle)^2}\\
&\le \sqrt{\left(1+\frac{m-1}{d}-\sum\limits_{k=1}^{m}\sum\limits_{i=0}^{d-1}\langle i_k|\rho^A|i_k\rangle^2\right)\left(1+\frac{m-1}{d}-\sum\limits_{k=1}^{m}\sum\limits_{i=0}^{d-1}\langle i_k|\rho^B|i_k\rangle^2\right)}.
\end{align*}
In the second inequality, we have used Cauchy-Schwarz  inequality. The third inequality is due to
\begin{align*}
 &\sum\limits_{k=1}^m\sum\limits_{i=0}^{d-1} \langle i_k|\varrho|i_k\rangle^2\le 1+\frac{m-1}{d} \text{ for any pure state } \varrho \text{ in } \mathbb{C}^d\;\;\cite{Wu2009},\\
 &\sum\limits_{k=1}^m\sum\limits_{i=0}^{d-1} \sum\limits_{s,t=1}^rp_sp_t\langle i_k|\rho_s^A|i_k\rangle\langle i_k|\rho_t^A|i_k\rangle=\sum\limits_{k=1}^m\sum\limits_{i=0}^{d-1}\langle i_k|\rho^A|i_k\rangle^2.
\end{align*}
$\hfill\Box$

The following proposition shows that Theorem 1 is more efficient than the MUB-criteria.\\
\\
\textbf{Proposition 1.} Theorem 1 is stronger than the MUB-criteria.
\\
\\
\textbf{Proof.} For any state $\rho$ in $\mathbb{C}^d\otimes \mathbb{C}^d$, we only need to prove the inequality (\ref{emubs}) holds if the inequality (\ref{memubs}) holds. In fact, assume that (\ref{memubs}) holds. Then we can get \[
L_m(\rho)\ge \sum\limits_{k=1}^{m}\sum\limits_{i=0}^{d-1}(\langle i_k|\otimes \langle i_k|\rho|i_k\rangle\otimes|i_k\rangle-\langle i_k|\rho^A|i_k\rangle \langle i_k|\rho ^B|i_k\rangle),\]
which, from the inequality $a+b\ge 2\sqrt{ab}, a,b\ge 0,$ implies
\begin{align*}
M_{m}(\rho)&=\sum\limits_{k=1}^{m}\sum\limits_{i=0}^{d-1}\langle i_k|\otimes \langle i_k|\rho |i_k\rangle\otimes|i_k\rangle\le L_m(\rho)+\sum\limits_{k=1}^{m}\sum\limits_{i=0}^{d-1}\langle i_k|\rho^A|i_k\rangle \langle i_k|\rho ^B|i_k\rangle\\
&\le 1+\frac{m-1}{d}-\frac{1}{2}\sum\limits_{k=1}^{m}\sum\limits_{i=0}^{d-1}\left(\langle i_k|\rho^A|i_k\rangle^2+\langle i_k|\rho^B|i_k\rangle^2-2\langle i_k|\rho^A|i_k\rangle \langle i_k|\rho ^B|i_k\rangle\right)\\
&\le 1+\frac{m-1}{d}.
\end{align*}
Thus, the inequality (\ref{emubs}) holds. $\hfill \Box$\\

\emph{Entanglement detection via MUMs}. Let $\{\mathcal{P}^{(b)}\}_{b=1}^{d+1} $ and $\{\mathcal{Q}^{(b)}\}_{b=1}^{d+1} $ be two sets of $d+1$ MUMs on $\mathbb{C}^d$ with the same parameter $\kappa$, where $\mathcal{P}^{(b)}=\{P_n^{(b)}\}_{n=1}^d$, $\mathcal{Q}^{(b)}=\{Q_n^{(b)}\}_{n=1}^d, b=1,2,\cdots,d+1$. For any separable state $\rho$ in $\mathbb{C}^d\otimes \mathbb{C}^d$, Chen et al. \cite{Chen2014-1} showed that
\begin{equation*}
\label{Chen1}
T(\rho)=\sum\limits_{b=1}^{d+1}\sum\limits_{n=1}^{d}\text{Tr}\left(P_n^{(b)}\otimes Q_n^{(b)}\rho\right)\le 1+\kappa.
\end{equation*}
This criteria will be called MUM-criteria. We now provide a criteria based on MUMs.\\
\\
\textbf{Theorem 2.} Let $\{\mathcal{P}^{(b)}\}_{b=1}^{d+1} $ and $\{\mathcal{Q}^{(b)}\}_{b=1}^{d+1} $ be two sets of $d+1$ MUMs defined as above. If the state $\rho$ in $\mathbb{C}^d\otimes \mathbb{C}^d$ is separable, then
\begin{align}
S(\rho)=\sum\limits_{b=1}^{d+1}&\sum\limits_{n=1}^{d}\left|\text{Tr}\left(P_n^{(b)}\otimes Q_n^{(b)}(\rho-\rho^A\otimes\rho^B)\right)\right|\nonumber\\
&\label{th21}\le \sqrt{\left(1+\kappa-\sum\limits_{b=1}^{d+1}\sum\limits_{n=1}^{d}(\text{Tr}(P_n^{(b)}\rho^A))^2\right)\left(1+\kappa-\sum\limits_{b=1}^{d+1}\sum\limits_{n=1}^{d}(\text{Tr}(Q_n^{(b)}\rho^B))^2\right)}.
\end{align}
\textbf{Proof.} Since $\rho-\rho^A\otimes \rho^B$ can be written as in the form (\ref{Zhang}), we get
\begin{align*}
S(\rho)&\le \frac{1}{2}\sum\limits_{n=1}^d\sum\limits_{b=1}^{d+1}\sum\limits_{s,t=1}^r \left|\sqrt{p_sp_t}\text{Tr}(P_n^{(b)}(\rho_s^A-\rho_t^A))\right|\left|\sqrt{p_sp_t}\text{Tr}(Q_n^{(b)}(\rho_s^B-\rho_t^B))\right|\\
&\le \frac{1}{2}\sqrt{\sum\limits_{n=1}^d\sum\limits_{b=1}^{d+1}\sum\limits_{s,t=1}^rp_sp_t(\text{Tr}(P_n^{(b)}(\rho_s^A-\rho_t^A)))^2}\sqrt{\sum\limits_{n=1}^d\sum\limits_{b=1}^{d+1}\sum\limits_{s,t=1}^rp_sp_t(\text{Tr}(Q_n^{(b)}(\rho_s^B-\rho_t^B)))^2}\\
&=\sqrt{\left(1+\kappa-\sum\limits_{b=1}^{d+1}\sum\limits_{n=1}^{d}(\text{Tr}(P_n^{(b)}\rho^A))^2\right)\left(1+\kappa-\sum\limits_{b=1}^{d+1}\sum\limits_{n=1}^{d}(\text{Tr}(Q_n^{(b)}\rho^B))^2\right)},
\end{align*}
where we have used the Cauchy-Schwarz inequality, and $
\sum\nolimits_{b=1}^{d+1}\sum\nolimits_{n=1}^{d}(\text{Tr}(P_n^{(b)}\varrho))^2=1+\kappa \text{ for any pure state } \varrho \text{ in }\mathbb{C}^d  \;\;\cite{Kalev2014}.\hfill \Box$

Similar to the proof of Proposition 1, we can easily get that Theorem 2 is stronger than the MUM-criteria.
For $\kappa=1$, Theorem 2 reduces to the case $m=d+1$ of Theorem 1.
\\

\emph{Entanglement detection via GSIC-POVMs}. Let $\{P_i\}_{i=1}^{d^2}$ and $\{Q_i\}_{i=1}^{d^2}$ be any two sets of GSIC-POVMs with the same parameter $\alpha$. It was shown in \cite{Chen2014-2} that, for any separable state $\rho$ in $\mathbb{C}^d\otimes \mathbb{C}^d$, it holds
\[
J(\rho)=\sum\limits_{i=1}^{d^2}\text{Tr}(P_i\otimes Q_i\rho)\le \frac{\alpha d^2+1}{d(d+1)}.
\]
It will be said to be GSICM-criteria. We now provide the criteria due to GSIC-POVMs.
\\
\\
\textbf{Theorem 3.} Let $\{P_i\}_{i=1}^{d^2}$ and $\{Q_i\}_{i=1}^{d^2}$ be any two sets of GSIC-POVMs with the same parameter $\alpha$. If the state $\rho$ in $\mathbb{C}^d\otimes \mathbb{C}^d$ is separable, then
\begin{align}
R(\rho)=&\sum\limits_{i=1}^{d^2}|\text{Tr}(P_i\otimes Q_i(\rho-\rho^A\otimes \rho^B))|\nonumber\\
\label{th31}&\le \sqrt{\left(\frac{\alpha d^2+1}{d(d+1)}-\sum\limits_{i=1}^{d^2}(\text{Tr}(P_i\rho^A))^2\right)\left(\frac{\alpha d^2+1}{d(d+1)}-\sum\limits_{i=1}^{d^2}(\text{Tr}(Q_i\rho^B))^2\right)}.
\end{align}
\textbf{Proof.}
It is trivial by an analogous argument as in Theorem 2 and
$\sum\nolimits_{i=1}^{d^2}(\text{Tr}(Q_i\varrho))^2=\frac{\alpha d^2+1}{d(d+1)}$
for any pure state $\varrho$ in $\mathbb{C}^d$ \cite{Rastegin2013}. $\hfill\Box$\\

Similar to Proposition 1, it is easy to prove that Theorem 3 is stronger than the GSICM-criteria.


 In what follows, we shall provide some examples to illustrate the efficiency of the presented criteria in this paper.\\
\\
\textbf{Example 1}. Consider the $d$-dimensional Bell-diagonal state \cite{Baumgartner2007} used in \cite{Spengler2012,Chen2014-1,Chen2014-2}
\[
\rho_{Bell}=\sum\limits_{s,t=0}^{d-1} c_{st}|\psi_{st}\rangle \langle \psi_{st}|,
\]
where $c_{st}\ge 0,\sum\nolimits_{s,t=0}^{d-1}c_{st}=1,|\psi_{st}\rangle=(U_{st}\otimes I_d)|\phi\rangle,$ and the Weyl operators $U_{st}, s,t=0,\cdots,d-1$ are defined as $U_{st}=\sum\nolimits_{j=0}^{d-1}\sigma_d^{sj}|j\rangle \langle j\oplus t|$ with $\sigma_d=e^\frac{2\pi \sqrt{-1}}{d}$ and $j\oplus t$ denoting $j+t$ mod $d$. It is easy to get $\rho_{Bell}^A=\rho_{Bell}^B=\frac{1}{d}I_d$. We set $$c_{ab}=\max\limits_{0\le s,t\le d-1}\{c_{st}\},c_{fg}=\min\limits_{0\le s,t\le d-1}\{c_{st}\}.$$

Since Theorem 1 and MUB-criteria with $m=d+1$ can be regarded as special cases of Theorem 2 and MUM-criteria, respectively, we do not compare them here.

\emph{Comparison of Theorem 2 and MUM-criteria}. Clearly, $S(\rho_{Bell})$ can be written as
\begin{equation}
\label{srho}
S(\rho_{Bell})=\sum\limits_{b=1}^{d+1}\sum\limits_{n=1}^{d}\left|\frac{1}{d}\sum\limits_{s,t=0}^{d-1}c_{st}\sum\limits_{i,j=0}^{d-1}\text{Tr}(P_n^{(b)}U_{st}|i\rangle\langle j|U_{st}^\dag)\text{Tr}(Q_n^{(b)}|i\rangle\langle j|)-\frac{1}{d^2}\right|,
\end{equation}
and the right hand side of the inequality (\ref{th21}) is $1+\kappa-\frac{d+1}{d}$.

We now consider the following two cases.

(i) $c_{ab}\ge\frac{1}{\kappa d}$. In this case, we set $Q_n^{(b)}=\overline{U_{ab}^\dag {P_n^{(b)}} U_{ab}}$, where $\overline{\cdot} $ denotes the conjugation of the corresponding matrix. From (\ref{srho}), it is easy to get
\begin{equation*}
\label{srho1} S(\rho_{Bell})\ge \kappa c_{ab}(d+1)-\frac{d+1}{d}.
\end{equation*}
 From Theorem 2, $\rho_{Bell} $ is entangled if $\kappa c_{ab}(d+1)-\frac{d+1}{d}>1+\kappa-\frac{d+1}{d}$, i.e., $c_{ab}>\frac{\kappa+1}{\kappa(d+1)}$.

(ii) $c_{fg}<\frac{1}{\kappa d}$. Taking $Q_n^{(b)}=\overline{U_{fg}^\dag {P_n^{(b)}} U_{fg}}$ leads to
\begin{equation*}
\label{lrho}
S(\rho_{Bell})\ge \frac{d+1}{d}-c_{fg}\kappa(d+1).
\end{equation*}
From Theorem 2, it follows that $\rho_{Bell}$ is entangled if $\frac{d+1}{d}-c_{fg}\kappa(d+1)>1+\kappa-\frac{d+1}{d}$, i.e.,
$c_{fg}<\frac{d-d\kappa+2}{\kappa d(d+1)}$.

By (i) and (ii), we conclude that if \begin{equation}
\label{cond11}c_{ab}>\max\left\{\frac{1}{\kappa d},\frac{\kappa+1}{\kappa(d+1)}\right\}=\frac{\kappa+1}{\kappa(d+1)}
\end{equation}
or
\begin{equation}
 \label{cond12}
 c_{fg}<\min\left\{\frac{1}{\kappa d},\frac{d-d\kappa+2}{\kappa d(d+1)}\right\}=\frac{d-d\kappa+2}{\kappa d(d+1)}
 \end{equation}
holds, then  $\rho_{Bell}$ is entangled.  However, by similar methods, the MUM-criteria can only detect the entanglement for $c_{ab}>\frac{\kappa+1}{\kappa(d+1)}$ \cite{Chen2014-1}.

It is obvious that the values of $\kappa $ affect the performance of the entanglement detection of Theorem 2. Surprisingly, entanglement conditions (\ref{cond11}) and (\ref{cond12}) are more efficient when $\kappa$ gets larger and smaller, respectively. Thus, in
order to detect more entanglement of Bell-diagonal states, the balance for the values of $\kappa$ is necessary. Another interesting conclusion can be drawn that, if one of the coefficients $c_{st},s,t=0,\cdots,d-1$ is zero, that is $c_{fg}=0$, then $\rho_{Bell}$ must be entangled by (\ref{cond12}).

\emph{Comparison of Theorem 3 and GSICM-criteria.} From $\text{Tr}(P_k)=\text{Tr}(Q_k)=\frac{1}{d}$,  it follows that $R(\rho_{Bell})$ can be written as
\begin{equation*}
\label{rrho}
R(\rho_{Bell})=\sum\limits_{k=1}^{d^2}\left|\frac{1}{d}\sum\limits_{s,t=0}^{d-1}c_{st}\sum\limits_{i,j=0}^{d-1}\text{Tr}(P_kU_{st}|i\rangle\langle j|U_{st}^\dag)\text{Tr}(Q_k|i\rangle\langle j|)-\frac{1}{d^4}\right|,
\end{equation*}
and the right hand side of (\ref{th31}) is $\frac{\alpha d^2+1}{d(d+1)}-\frac{1}{d^2}$.

The following two cases are considered.

(i) $c_{ab}\ge \frac{1}{d^3 \alpha}$. Taking $Q_k=\overline{U_{ab}^\dag P_k U_{ab}}$ results in
\begin{equation*}
\label{rrhol} R(\rho_{Bell})\ge d\alpha c_{ab}-\frac{1}{d^2}.
\end{equation*}
From Theorem 3, $\rho_{Bell} $ is entangled if $d\alpha c_{ab}-\frac{1}{d^2}>\frac{\alpha d^2+1}{d(d+1)}-\frac{1}{d^2}$, i.e.,
 $c_{ab}>\frac{\alpha d^2+1}{\alpha d^2(d+1)}$.

(ii) $c_{fg}<\frac{1}{d^3 \alpha}$. Taking $Q_k=\overline{U_{fg}^\dag P_k U_{fg}}$ we get
\begin{equation*}
\label{rrhou} R(\rho_{Bell})\ge \frac{1}{d^2}-d\alpha c_{fg}.
\end{equation*}
By Theorem 3, $\rho_{Bell} $ is entangled if $\frac{1}{d^2}-d\alpha c_{fg}>\frac{\alpha d^2+1}{d(d+1)}-\frac{1}{d^2}$, i.e.,  $c_{fg}<\frac{d-d^3\alpha+2}{d^3\alpha(d+1)}$.

Due to (i) and (ii), we conclude that, if
\[
c_{ab}>\max\left\{ \frac{1}{d^3 \alpha},\frac{\alpha d^2+1}{\alpha d^2(d+1)}\right\}=\frac{\alpha d^2+1}{\alpha d^2(d+1)}
\]
or
\[
c_{fg}<\min\left\{\frac{1}{d^3 \alpha},\frac{d-d^3\alpha+2}{d^3\alpha(d+1)}\right\}=\frac{d-d^3\alpha+2}{d^3\alpha(d+1)}
\]
holds, then $\rho_{Bell}$ is entangled. However, by similar methods, the GSICM-criteria can only detect the entanglement for $c_{ab}>\frac{\alpha d^2+1}{\alpha d^2(d+1)}$ \cite{Chen2014-2}.

Similar to the parameter $\kappa$ in Theorem 2, the parameter $\alpha$ also plays an important role in the entanglement detection of Theorem 3.
\\
\\
\textbf{Example 2.} Consider the $d$-dimensional Werner states \cite{Werner1989}
\[
\rho_W=\frac{1}{d^3-d} \left((d-g)I_{d^2}+(dg-1)\eta\right),
\]
where $-1\le g\le 1$, the ``flip" or ``swap" operator $\eta$ can be represented as $\eta=\sum\nolimits_{i,j=0}^{d-1}|ij\rangle\langle ji|$, and $\rho_W$ is entangled if and only if $-1\le g<0$.

\emph{Comparison of Theorem 2 and MUM-criteria.} Taking $Q_n^{(b)}=P_n^{(b)}$ and simple computation yield
\[
T(\rho_W)=\frac{1}{d-1}\left((d-g)+\kappa(dg-1)\right).
\]
Since $T(\rho_W)>1+\kappa$ if and only if $g>1$, MUM-criteria can not detect any entanglement of Werner states.

We now use Theorem 2 to detect the entanglement of Werner states. The left and right hand sides of (\ref{th21}) are, respectively,
\[
S(\rho_W)=\left|\frac{1}{d-1}\left((d-g)+\kappa(dg-1)\right)-\frac{d+1}{d}\right| \text{ and } 1+\kappa-\frac{d+1}{d}.
\]
Since $S(\rho_W)>1+\kappa-\frac{d+1}{d}$ if and only if $g>1$ or $g<\frac{2}{d}-1,$ from Theorem 2 we get $\rho_W$ is entangled for $-1 \le g<\frac{2}{d}-1$. Thus, Theorem 2 can detect completely any entanglement of the 2-dimensional Werner states.

\emph{Comparison of Theorem 3 and GSICM-criteria.} Setting $Q_k=P_k$, we get
\[
J(\rho_W)=\frac{d-g}{d^3-d}+\frac{d\alpha(dg-1)}{d^2-1}.
\]
Since $J(\rho_W)>\frac{\alpha d^2+1}{d(d+1)} $ if and only if $g>1$, similar to MUM-criteria, GSICM-criteria can not detect any entanglement in Werner states.

We now apply Theorem 3. The left and right hand sides of (\ref{th31}) are, respectively,
\[
R(\rho_W)=\left|\frac{d-g}{d^3-d}+\frac{d\alpha(dg-1)}{d^2-1}-\frac{1}{d^2}\right| \text{ and } \frac{\alpha d^2+1}{d(d+1)}-\frac{1}{d^2}.
\]
Since $R(\rho_W)> \frac{\alpha d^2+1}{d(d+1)}-\frac{1}{d^2}$ if and only if $g>1$ or $g<\frac{2}{d}-1$, Theorem 3 can detect the entanglement of $\rho_W$ for $-1\le g< \frac{2}{d}-1$.

As a conclusion, the detection ability of Theorem 3 is the same as that of Theorem 2 for $d$-dimensional Werner states. Both of them are more efficient than the MUM-criteria and GSICM-criteria.\\
\\
\textbf{Example 3}. The following well-known $3\times 3$ bound entangled state was given by Horodecki \cite{Horodecki1997}:
\[\rho=\frac{1}{8a+1}\left( {\begin{array}{*{20}{c}}
   {a} & {0} & {0} & {0} & {a} & {0} & {0} & {0} & {a}  \\
   {0} & {a} & {0} & {0} & {0} & {0} & {0} & {0} & {0}  \\
   {0} & {0} & {a} & {0} & {0} & {0} & {0} & {0} & {0}  \\
   {0} & {0} & {0} & {a} & {0} & {0} & {0} & {0} & {0}  \\
   {a} & {0} & {0} & {0} & {a} & {0} & {0} & {0} & {a}  \\
   {0} & {0} & {0} & {0} & {0} & {a} & {0} & {0} & {0}  \\
   {0} & {0} & {0} & {0} & {0} & {0} & {\frac{1+a}{2}} & {0} & {\frac{\sqrt{1-a^2}}{2}}  \\
   {0} & {0} & {0} & {0} & {0} & {0} & {0} & {a} & {0}  \\
   {a} & {0} & {0} & {0} & {a} & {0} & {\frac{\sqrt{1-a^2}}{2}} & {0} & {\frac{1+a}{2}}  \\
\end{array}} \right),\]
where $0<a<1$. We consider the mixture of this state with the maximally entangled state
$|\phi\rangle =\frac{1}{\sqrt{3}}\sum\nolimits_{i=0}^{2}|ii\rangle$:
$$
\rho_{mix}=p\rho +(1-p)|\phi\rangle\langle \phi|,
$$
 where $0\le p\le 1$. The complete sets of MUMs and GSIC-POVMs are constructed from the generalized Gell-Mann operators; see \cite{Kalev2014} and \cite{Kalev2013} for the detailed constructions. If we take $$Q_n^{(b)}=\overline{P_n^{(b)}},Q_k=\overline{P_k},$$ then from MATLAB computations we find that the same entangled conditions can be obtained by MUM-criteria and GSICM-criteria (respectively, Theorems 1 and 2) for a fixed value of $a$. The numerical results with different values of $a$ are displayed in Table 1. It can be seen that Theorems 1-2 are more efficient than the MUM-criteria and GSICM-criteria. Nevertheless, the difference between them becomes smaller as $a$ gets larger.

\begin{table}[htbp]
\centering \begin{tabular} {c|c|c}\hline
\raisebox{-1.50ex}[0cm][0cm]{$a=0.1250$}& MUM or GSICM-criteria& $0\le p\le 0.9015$ \\ \cline{2-3}
 &
Theorem 1 or 2& $0\le p\le 0.9456$
 \\ \hline \raisebox{-1.50ex}[0cm][0cm]{$a=0.3750$}& MUM or GSICM-criteria& $0\le p\le 0.9625$
  \\ \cline{2-3}
 &
Theorem 1 or 2& $0\le p\le 0.9713$ \\ \hline \raisebox{-1.50ex}[0cm][0cm]{$a=0.6250$}& MUM or GSICM-criteria&$0\le p\le 0.9847$ \\ \cline{2-3}
&
Theorem 1 or 2& $0\le p\le 0.9872$ \\ \hline
\raisebox{-1.50ex}[0cm][0cm]{$a=0.8750$}& MUM or GSICM-criteria&$0\le p\le 0.9962$ \\ \cline{2-3}
&
Theorem 1 or 2& $0\le p\le 0.9966$ \\ \hline
\end{tabular} \caption{\emph{Entanglement conditions of $\rho_{mix}$ from MUM-criteria, GSICM-criteria and Theorems 1-2}}
\label{tab5}
\end{table}

\section{Conclusion}

In this paper, MUBs, MUMs and GSIC-POVMs have been used to study the entanglement detection of the bipartite quantum systems. Based on $\rho-\rho^A\otimes \rho^B$, we have presented three separability criteria, i.e., Theorems 1-3, which, by strict proofs, are more efficient than the detection methods introduced in \cite{Spengler2012,Chen2014-1,Chen2014-2}, respectively. By taking Bell-diagonal states, Werner states, the mixture of Horodecki's $3\times 3$ bound entangle state and the maximally entangled state as examples, we have shown that the performance of the presented criteria depends on not only the values of involved parameters $\kappa$ and $\alpha$, but also the choice of the complete set of measurements.

Compared with some criteria including PPT criteria and realignment criteria, the criteria given by Theorems 1-3 are relatively easy to be implemented experimentally, since only some local measurements are used. The relations between the presented criteria and other detection methods such as PPT criteria, realignment criteria and correlation matrix criteria need to be studied in the future. How to construct new efficient separability criteria by using measurements is also an interesting problem.


\section*{\bf Acknowledgments}
 This work is supported by the
NSFC 11105226; the Fundamental Research Funds for the
Central Universities No.12CX04079A, No.24720122013; Research Award
Fund for outstanding young scientists of Shandong Province
No.BS2012DX045. The authors are grateful to the referee and the editor for their invaluable comments and
suggestions, and would like to thank Prof. Chi-Kwong Li at Department of Mathematics, College of William and Mary for helpful discussions. The research was done while the first author was visiting the College of William and Mary during the academic year 2013-14 under the support of China Scholarship Council.

\end{document}